# Site-specific spin crossover in $Fe_2TiO_4$ post-spinel under high pressures to near a megabar

W. M. Xu,[1,*] G. R. Hearne,[2] S. Layek,[1] D. Levy,[1] J-P. Itié,[3] M. P. Pasternak,[1] G. Kh. Rozenberg[1] and E. Greenberg[1,4]

[1]*School of Physics and Astronomy, Tel Aviv University, 69978 Tel Aviv, Israel*

[2]*Department of Physics, University of Johannesburg, P.O. Box 524 , Auckland Park, 2006, Johannesburg, South Africa*

[3]*Synchrotron SOLEIL, L'Orme des Merisiers, Saint-Aubin, BP 48, 91192 Gif-sur-Yvette Cedex, France*

[4] *Center for Advanced Radiation Sources, University of Chicago, Argonne, IL 60439, USA*

## Abstract

X-ray diffraction studies to ~90 GPa at room temperature show that $Fe_2TiO_4$ ferrous inverse spinel undergoes the following sequence of structural transitions :

$$\text{cubic } (Fd\bar{3}m) \xrightarrow{\sim 8\text{ GPa}} \text{tetragonal } (I4_1/amd) \xrightarrow{\sim 16\text{ GPa}} \text{orthorhombic}(Cmcm) \xrightarrow{\sim 55\text{ GPa}} \text{orthorhombic}(Pmma),$$

at the indicated onset transition pressures. Within the *Cmcm* phase, site-specific spin crossover is initiated and involves only highly distorted octahedral sites constituting ~25% of all Fe locations. This is manifest as a steeper volume decrease of $\Delta V/V_0 \sim 3.5\%$ beyond ~40 GPa and an emergent diamagnetic component discerned in $^{57}$Fe Mössbauer spectroscopy at variable cryogenic temperatures. A subsequent *Cmcm* → *Pmma* Fe/Ti disorder-order reconfiguration is facilitated at 6-fold coordinated (octahedral) sites. The rest of the high-spin Fe in 6-fold and 8-fold coordinated sites (~75% abundance) in the *Pmma* phase exhibit average saturation internal magnetic fields of $H_{hf}$ ~ 42 T to ~90 GPa, typical of spin-only (orbitally quenched) Fermi-contact values. By contrast average $H_{hf}$ ~ 20 T values, signifying unquenched orbital moments, occur below the 40–45 GPa spin-crossover initiation regime in the *Cmcm* phase. Therefore site-specific spin crossover invokes a cooperative lattice response and polyhedral distortions at the rest of the high-spin Fe sites, translating to 3*d* level (sub-band) changes and consequential orbital moment quenching. Near ~90 GPa $Fe_2TiO_4$ is a partially spin-converted chemically ordered *Pmma* post-spinel structure having a persistent charge gap of ~100 meV. Despite structural symmetry changes, partial spin crossover and lattice compressibility resulting in a ~33% total reduction in unit-cell volume and corresponding 3*d* bandwidth broadening, strong electron correlations persist at high densification.

* Author to whom all correspondence should be addressed : xuw@post.tau.ac.il

## I. Introduction

Ulvöspinel, $Fe_2TiO_4$, is a II-IV inverse spinel which contains ferrous ions which occupy both tetrahedral (*A*) sites and octahedral (*B*) sites. $Fe^{2+}$ and $Ti^{4+}$ cations are disordered on the *B* sub-lattice and the spinel formulation may be rendered as follows $Fe^{2+}(Fe^{2+}Ti^{4+})O_4$ [1, 2], where the cations inside the parentheses occupy octahedral *B* sites and those outside occupy tetrahedral *A* sites. At ambient conditions $Fe_2TiO_4$ is cubic (SG : $Fd\overline{3}m$). It transforms to the tetragonal phase (SG: $I4_1/amd$) below 163K due to a Jahn-Teller distortion associated with $Fe^{2+}$ at the tetrahedral site [1]. The onset of antiferromagnetism occurs at $T_N$ ~ 136 K from the opposing spin alignments on *A* and *B* sub-lattices. Appreciable distortions occur along crystallographic directions in the magnetically ordered state. Both this giant magnetostriction and increases in magnetocrystalline anisotropy are attributed to an important role of tetrahedral $Fe^{2+}$ and associated Jahn-Teller effects [3-6].

Of no less importance or interest is that ulvöspinel is an end-member of the important titanomagnetite solid solution between $Fe_3O_4$ (magnetite) and $Fe_2TiO_4$. Compositions of titanomagnetite are the dominant carrier of magnetic remanence in nature and have central importance to paleomagnetic, rock magnetic, and mineral magnetic studies [7, 8]. Titanomagnetite solid solutions may be formulated as $Fe_{3-x}Ti_xO_4$, where $x$ defines the proportion of ulvöspinel in the solid solution. The distribution of $Fe^{2+}$, $Fe^{3+}$ and $Ti^{4+}$ cations is of crucial importance in determining the magnetic properties pertinent to rock magnetism. Whereas antiferromagnetic $Fe_2TiO_4$ has spin compensating octahedral and tetrahedral sub-lattices, magnetite $Fe^{3+}(Fe^{2+}Fe^{3+})O_4$ in turn is a well known ferrimagnet with $T_C$ ~ 860 K. Moreover the magnetostriction and magnetocrysalline anisotropy mentioned above for $Fe_2TiO_4$ also occur in the solid solution series $Fe_{3-x}Ti_xO_4$ and the extent of these magnetic responses are positively dependent on the ulvöspinel content [9, 10].

Extensive high pressure studies have already been conducted on the one end-member, magnetite $Fe_3O_4$ , see Glazyrin *et al.* [11] for studies to ~25 GPa and Xu *et al.* [12] for investigations to higher pressures exceeding 50 GPa, and references cited in those publications. Strong on-site repulsion effects associated with the partially filled narrow 3*d* bands render insulating behavior to mixed-valence magnetite $Fe^{3+}(Fe^{2+}Fe^{3+})O_4$ at ambient pressure. These strong electron correlation effects and non-metallic behavior seem to persist to very high pressures exceeding ~50 GPa [13]. Beyond this pressure only annealing at high temperatures of 1050 K [14], or pressurizing to much higher pressures of ~120 GPa at room temperature yields a positive temperature coefficient of resistance and resistivity values typical of a poor metal in the vicinity of room temperature [12]. There is controversy about magnetic-electronic aspects of the compound in the low pressure regime where there are claims and counter-claims for an electronic change from x-ray magnetic circular dichroism and $^{57}$Fe MS investigations, respectively [11, 15].

Here we conduct a similar comparative investigation of the other end member $Fe^{2+}(Fe^{2+}Ti^{4+})O_4$. By using a combination of variable cryogenic temperatures $^{57}$Fe Mössbauer-effect spectroscopy (MS) and electrical-transport ( resistance) probes as well as complementary x-ray diffraction (XRD) structural studies at in-situ pressures to ~90 GPa, we provide insight into the interplay between the magnetic-electronic aspects and lattice-structural responses at varying degrees of high densities.

It is also anticipated that this will serve as important reference information when interpolating to members of the solid-solution series of pertinence to rock magnetism and the geosciences. In addition, spinels represent an abundant structure-type in both the lower crust and the upper mantle. They are known to exhibit a pressure-induced phase transformation to orthorhombic post-spinel $CaFe_2O_4$- , $CaMn_2O_4$- and $CaTi_2O_4$-type structures; for a review see the reference of Erandonea [16]. There are hardly any magnetic-electronic investigations of these post-spinel phases in comparison to the extensive structural investigations already published [16].

Previous high pressure XRD structural studies have demonstrated that $Fe_2TiO_4$ undergoes a series of structural phase transitions from cubic ( $Fd\bar{3}m$ ) to tetragonal ($I4_1/amd$) at ~9 GPa, and then to an orthorhombic post-spinel structure (*Cmcm*) at ~16 GPa [1, 17]. The tetrahedral sites of the spinel evolve to 8-fold coordinated sites in the post-spinel structure [18], in addition to structurally adjusted 6-fold coordinated sites occurring in which the Fe/Ti remains chemically disordered. Yamanka *et al.* [1, 17] have also shown that in a number of $Fe_{3-x}Ti_xO_4$ compositions the transition pressure to an orthorhombic post-spinel phase has a nearly linear dependence on $x$ at room temperature. It varies from 25 GPa for $x = 0$ (magnetite) to ~16 GPa for $x = 1$ in the other end member, ulvöspinel. In particular in $Fe_2TiO_4$ another high pressure polymorph was deduced to occur above 50 GPa. Refinements of XRD data in the range 50–60 GPa in terms of *Pmma* symmetry, corresponding to Fe/Ti chemical order in 6-fold coordinated sites, yield the highest reliability factors [17]. This represents a non-isomorphic sub-group of the *Cmcm* space group. Later Raman pressure studies to ~57 GPa [19], seem to corroborate this sequence of structural transitions. Yamanaka *et al.* [17] also present Fe $K\beta$ x-ray emission spectroscopy (XES) results for $Fe_2TiO_4$. They contend therein that spin crossover to diamagnetic low-spin Fe initiates as low as ~14 GPa at room temperature and progresses to completion by ~30 GPa. This would have important geo-magnetic implications given the role of the $Fe_3O_4$-$Fe_2TiO_4$ solid solution series in this regard. The XES experiments do not reveal a detailed picture of the crystal-chemistry changes associated with the purported spin crossover and its relation to the structural transitions mentioned above. For example, does the spin pairing occur at all of the 6-fold coordinated Fe sites ? A much more suitable probe of this detail is obtained by variable temperature $^{57}$Fe MS, which is a direct probe of the Fe magnetic-electronic state even under in-situ extreme pressure-temperature conditions. In this regard exceedingly important and unique information may be discerned, in that spin and oxidation states and their distribution and abundances in the lattice structure are obtained from the deconvolution of sub-components in a spectrum [20]. We implement this to ~90 GPa, well beyond previous high pressure MS studies on ulvöspinel by Wu *et al.* [21] limited to 24 GPa at room temperature. In addition we extend the XRD measurements and analysis to ~90 GPa, also to well beyond the previous investigations of Yamanaka *et al.* [1, 22]. Our complementary electrical resistance measurements complete the picture of the pressure evolution of strong electron correlations in the various structural phases.

## II. Experimental

The $Fe_2TiO_4$ sample is the same as that used in the previous study by Wu *et al.* [21]. This involved a solid-state reaction of stoichiometric quantities of $Fe_2O_3$ and $TiO_2$ at a predetermined oxygen fugacity using an appropriate CO-$CO_2$ mixture. The $Fe_2O_3$ precursor material was enriched to ~63% $^{57}$Fe by thorough mixing of near fully enriched and natural abundance oxides. The cubic structure and magnetic properties were compatible with previous reports of phase pure samples at ambient pressure [1, 2], as confirmed by conventional powder XRD and also by variable cryogenic temperature MS in our laboratory.

*$^{57}$Fe Mössbauer-effect* studies were carried out using a $^{57}$Co(Rh) 10-mCi point source in the 5–300 K temperature range involving a top-loading liquid-helium cryostat [23]. The typical collection time of a spectrum was ~24 hours. Spectra at the highest pressures near ~90 GPa at variable cryogenic temperatures, involving much shorter data collection times, were recorded at the European Synchrotron Radiation Facility (ESRF, Grenoble–France) using synchrotron Mössbauer source methodology [24]. Spectra were analyzed using appropriate fitting programs from which the hyperfine interaction parameters and the corresponding relative abundances of the spectral components were derived [25, 26]. The isomer shift (IS) in the present work is calibrated relative to an α-Fe foil at ambient conditions. The *Tel-Aviv University* (TAU) miniature piston-cylinder DAC was used [27], with anvils having 300-μm diameter culets for pressures to ~60 GPa and 200-μm diameter culets for $P > 60$ GPa. Samples were loaded into 100 to 150-μm diameter cavities drilled in a rhenium gasket for MS studies. This also served as an effective collimator for the 14.4 keV γ-rays. Liquid Argon was used as a pressurizing medium [28, 29]. A few ruby chips were included in the sample cavity for pressure measurements by way of the ruby *R1* fluorescence pressure marker.

*Powder XRD* measurements were carried out at room temperature in angle-dispersive mode with a wavelength of λ = 0.3738 Å at the *Pression Structure Imagerie par Contraste à Haute Énergie* (PSICHÉ) beamline of Synchrotron Soleil (Paris). Pressurization was by means of a TAU piston-cylinder DAC with anvils having 200-μm diameter culets. The sample with a few ruby chips was loaded into a 100-μm diameter cavity drilled in a rhenium gasket pre-indented down to a final thickness of ~15 μm and Ne or He was used as the pressure transmitting medium. Diffraction images were collected using a MAR345 image plate detector and integrated using the FIT2D [30, 31] and DIOPTAS software [32] . Powder diffraction patterns were analyzed using the GSAS-II software [33] to extract the unit-cell parameters.

The intensities of the diffraction peaks are affected by instrumental and grain-size issues (diamond x-ray absorption and low statistics in random distribution of the sample crystallites). Therefore the Rietveld refinement of the powder diffraction patterns did not result in a good enough fit. Hence diffraction patterns were analyzed by using the whole profile fitting (Pawley) method [34]. Diffraction spectra mainly exhibit peaks from the $Fe_2TiO_4$ sample, as well as minor peak contributions from the gasket (Re) and the pressure medium (Ne). These phases were considered in the refinements. The reliability factor $R_{wp}$ obtained in the refinement of each of the powder diffraction patterns is ~1%.

*Resistance measurements* in the TAU piston-cylinder DAC involved sample loaded into ~100-μm cavities drilled in rhenium gaskets insulated with a layer of $Al_2O_3$-NaCl (3:1 *wt. %*) mixed with epoxy. Conductive epoxy was used to connect exterior conducting Cu wires to microscopic triangular cut Pt foils. The latter served as electrodes in a DC four-probe configuration, leading from the pavilion of the anvil to near the center of the culet so as to overlap with the sample upon closure of the DAC. Six of such electrodes were configured on the culet to permit measurements in various four-probe arrangements at a given pressure. This built-in redundancy affords a means of checking the effects of any pressure distribution in the regions between the leads and provides alternatives in case of any lead breakages in the compression sequence. A few ruby fragments for pressure determination were located in the region between the Pt electrode tips overlapping the sample. No pressure transmitting medium was used, but pressure is effectively transmitted to the sample upon compression by way of the surrounding insulation. Pressure gradients are expected to be small in the distances (20–30 μm) between the tips of the Pt electrodes across which voltage measurements are made. For variable low temperature measurements the DAC was placed on a probe connected to a "dip-stick" stepper-motor assembly, which by computer control slowly changed the height of the DAC above the cryogen level in a liquid nitrogen or helium Dewar. The temperature was monitored by a Lakeshore Si (DT-421-HR) diode in proximity to the DAC.

The calibration scales mentioned in ref. [35] were used for pressure determination from the ruby fluorescence measurements. The error in the pressure determination is 5–10% of the reported average pressure from the ruby fluorescence measurements in the case of the MS and resistance experiments. In the case of the XRD measurements, pressure was determined from the ruby fluorescence spectra up to ~51 GPa. The Ne equation of state was used to ascertain pressures in the range 10–85 GPa [36]. Both methods give rather similar pressure values; the difference not exceeding 0.5 GPa. The error in the pressure determination is about 2% of the reported average pressure.

## III. Results and discussion

The magnetic-electronic behavior up to ~90 GPa is summarized in the evolution of representative Mössbauer spectra at room temperature (RT) and lowest cryogenic temperatures (LT) in Fig. 1.

In fitting the Mössbauer spectra in this range we first consider the spectrum and analysis of the spinel $Fe^{2+}(Fe^{2+}Ti^{4+})O_4$ at ambient pressure. Fe/Ti is disordered on the octahedral *B* sites and this leads to a distribution of Fe local environments (next nearest neighbor variations). The tetrahedral *A* sites share some corners with the *B* sites and therefore the associated tetrahedral Fe also experience a distribution of next nearest neighbor environments. Therefore the hyperfine interaction (HI) parameters most directly affected by the local lattice environment and the atomic disorder, quadrupole splitting/shift QS and magnetic hyperfine field $H_{hf}$, will have a distribution of values [25]. The spectrum at room temperature has a broadened doublet profile with poorly resolved shoulder features in the inner parts. This hints at two strongly overlapping QS contributions from the *A* and *B* sites which are not well resolved. This has been rationalized from the orbital level scheme and how the degeneracies are lifted due to the distortions at the *A* and *B* sites. The minority spin (↓) occupation of the lowest lying 3*d* orbital levels has the most significant effect on the resultant electric field gradient (EFG) and consequently determines the QS [37]. This similar occupation of the lowest lying 3*d* orbital singlet for *A* and *B* sites leads to comparable QS values and hence the strong overlap of components from these sites [38, 39]. Moreover there is a distribution of QS values from the Fe/Ti disorder effects which is expected to result in multiple *B*-site local environments and consequentially multiple *A*-site local environments as well. All of this may be manifested as appreciable line broadening of discrete sub-spectra fitted to the spectral profile. Similar unresolved features occur in the complex magnetic spectra at $T \ll T_N$, attributable to both a distribution of $H_{hf}$ values from the varied magnetic superexchange contributions as well as QS distribution associated with the Fe/Ti disorder on *B*-sites. Because of the strong overlap of *A*- and *B*-site contributions it is difficult to determine what the true distributions are for the QS and $H_{hf}$ parameters. Nakamura and Fuwa [2] have rather used a combination of two broadened Lorentzian sub-spectra to represent the distributions associated with the *A* sites and similarly associate two components with the *B* sites. This four-component fitting is the minimum number of Lorentzian profiles required to render a good fit to the overall spectral envelope, over an extended temperature range 300–16 K. In each case of both *A* and *B* sites the two broadened components are supposed to represent two different degrees of partial order in the unit cells, and not a completely random arrangement of Fe/Ti on the *B* sub-lattice [2].

We have adopted a similar minimal number of broadened discrete components model to analyze our temperature dependent spectra measured at high pressure. Spectra taken at high pressure in the DAC also have extrinsic sources of broadening from the customized point source [23], as well as from pressure gradients and the associated stress distribution in the pressurized cavity. So the true distributions associated with the HI parameters would be even more difficult to determine unambiguously. The IS , QS and $H_{hf}$ values derived from the fitting with broadened Lorentzian components, then represent the averages of the associated parameter distributions and are quite insensitive to the degree of line broadening (be it intrinsic or from extrinsic effects [40]). This is expected to ensure reliable qualitative interpretation of the data. As a check, selected spectra fitted with the full transmission integral yielded similar HI parameters and relative abundances of sub-components to that obtained in fitting with Lorentzian profiles.

Hyperfine interaction parameters are constrained, or starting values chosen, to ensure consistency between RT and LT analyses. Linewidth broadening is permitted to artificially account for the local environment distribution and thickness effects from the isotopic enrichment [40]. The pressure evolution of these parameters are depicted in Fig. 2, as derived from typical fits shown in Fig. 1. Three pressure regimes are delineated, partly based on previous structural data of Yamanaka *et al.* [1, 17], and these are discussed in detail in the following sub-sections. The results and analysis of Mössbauer data are correlated to the pressure dependences of our XRD structural data in Figs. 3 and 4, as well as resistance data (Fig. 5) in these regimes.

### A. Cubic and tetragonal phases, ambient pressure to ~16 GPa

X-ray investigations from ambient pressure to 8 GPa show the structure to be cubic spinel at room temperature. From thereon up to 16 GPa it is tetragonally distorted [1].

Fig. 1 shows RT and LT Mössbauer spectra at 1 and 14 GPa representative of this low pressure regime. The profile of the spectra at RT have narrowed and become more symmetrical compared to the spectrum at ambient pressure. Using the linewidth (0.4 mm $s^{-1}$) fitted at ambient pressure, the RT spectra above 1 GPa could be satisfactorily fitted with three quadrupole doublets with an intensity ratio 1:0.5:0.5, supposed to represent high-spin (HS) Fe in the *A* sites (HS(IV)) and two categories of *B* sites (HS(VI-1), HS(VI-2)), respectively. This provided the best fit to both LT and RT spectra and is a slight variant on the four component model of ambient pressure, in that the two suites representing tetrahedral *A* sites are now merged whereas the two suites associated with the octahedral *B* sites are retained. This change is a result of the sharpening of the spectrum. The change occurs mainly with the *A*-site QS components likely because these also have contributions from the dynamical Jahn-teller effect in the cubic structural phase. These are probably very sensitive to pressurization, similar to the sensitivity seen in varying the temperature [20]. A plot of the pressure dependence of the HI parameters is depicted in Fig. 2.

The three components have the same isomer (centroid) shift (IS) but different QS values. The component with largest QS ~ 2.5 mm $s^{-1}$ has a large average field value of $H_{hf}$ ~ 40 T from its saturation magnetization at 50 K. We attribute this to one category of 6-fold coordinated (octahedral) sites, designated HS(VI-2). The other two sub-spectra have low average field values $H_{hf}$ ~ 20 T but with different QS values of 1 mm $s^{-1}$ and 1.5 mm $s^{-1}$, attributed to a second category of octahedral sites designated HS(VI-1) and to the tetrahedral sites designated HS(IV), respectively. It should be noted that the quoted $H_{hf}$ values must be interpreted as average values of the true distributions which are represented here by broadened Lorentzian profiles. In these magnetic spectra at LT the overall spectral profile dictates that both *B* sites have negative QS values and the *A* site has a positive QS, as explained in ref. [2].

It may also be noted in Fig. 2 that from 1 GPa to ~8 GPa in the cubic phase at RT that the QS of the sites have a much stronger pressure dependence than when the compound becomes tetragonal as pressure rises above 8 GPa. This is another indicator of the sensitivity of dynamical Jahn-Teller distortions and its associated QS contribution to applied pressure and stresses in the sample, as mentioned earlier. This also marks a boundary where the cubic-to-tetragonal transition temperature has risen from $T_{J\text{-}T}$ ~ 163 K at ambient pressure up to room temperature by ~8 GPa.

As in the interpretation of spectra at ambient pressure the two categories of HS(VI) octahedral sites in this pressure regime are ascribed to two levels of partial Fe/Ti order in the *B* sub-lattice [2]. In Fig.2 the large field $H_{hf}$ ~ 40 T of HS(VI-2) is comparable to the spin-only value obtained from the Fermi contact interaction term $H_C = +(const.)\mu_B\left[\rho_s^{\uparrow}(0) - \rho_s^{\downarrow}(0)\right]$ involving the net spin-up and spin-down *s*-electron density at the nucleus. The $H_{hf}$ ~ 20 T values of HS(IV) and HS(VI-1) components are much smaller than typical ferrous values of $H_C$ ~ 44 T. This is suggestive of an opposing unquenched orbital moment term $H_L = -2\mu_B\left\langle r^{-3}\right\rangle\left\langle l_Z\right\rangle$ contributing to $H_{hf} = H_C + H_L$ at these sites [41-43]. According to Nakamura and Fuwa [2] these are also sites where the partial Fe/Ti order in the unit cells of the *B* sub-lattice is distinct from that of HS(VI-2).

X-ray pressure results of our sample are consistent with that of Yamanaka *et al.* [1]. Diffraction patterns at RT in the range 8–16 GPa may be refined in terms of a tetragonal phase (SG : *I4/amd*), see Figs. 3 and 4. As the tetragonal phase emerges at 8 GPa the weak monotonic decrease of the resistance at RT in the cubic phase changes to a much stronger negative pressure dependence as the tetragonal phase is pressurized, see Fig. 5. This results in a decrease in resistance by more than two orders of magnitude in the range 8–20 GPa, after which there is a change back to a much weaker pressure dependence again for $P > 20$ GPa. The data is well represented by activated transport processes $R = R_0 exp(E_a/(k_B T))$, where $E_a$ is the activation energy, as seen by the linearized data sets $ln(R)$ vs $1/T$ in the inset of Fig. 5. The changes in $E_a$ values in the inset mirror the appreciable changes in pressure dependences of the $R$(300 K) data in the main panel. We consider this corroboration of the ~8 GPa and ~16 GPa phase transition boundaries identified in the XRD data. Any difference in transition pressures identified by the two techniques is attributed to the degree of hydrostaticity from the pressure transmitting medium used in the pressurized cavity, see experimental section-II.

### B. *Cmcm* post-spinel phase in the range 20–55 GPa, initiation of site-specific spin crossover and orbital moment quenching

The emergence of a new structural phase beyond 16 GPa is associated with a substantial decrease in unit-cell volume, see Fig. 4. The x-ray diffraction results of Yamanaka *et al.* [17] and our data and analysis in Figs. 3 and 4, show that above ~20 GPa $Fe_2TiO_4$ transitions to the orthorhombic post-spinel $CaTi_2O_4$-type structure with space group *Cmcm*. There is also a change to a much lower compressibility with a substantial increase in bulk modulus from $K_0$ ~ 144 GPa to 319 GPa.

In this post-spinel structure $Fe^{2+}$ occupies so-called M1 (8-fold coordinated) sites, and $Fe^{2+}$ and $Ti^{4+}$ occupy M2 (6-fold coordinated) sites with occupancy ratio 1:1:1 [17], see Fig. 6. In this pressure regime the spectral envelope of Mössbauer data at RT become broader and more asymmetrical. Yet the low temperature spectra remain somewhat similar to the spinel phase in that both large and low average $H_{hf}$ distributions can be fitted, Fig 1, arising from the Fe/Ti partial order proposed in the 6-fold coordinated sites [2]. The Mössbauer spectra at RT are best fit with three components as in the spinel phase, namely one M1 (designated HS(VIII)) and two M2 sites, designated HS(VI-1) and HS(VI-2), conforming to a ratio of 1:0.5:0.5 for HS(VIII):HS(VI-1):HS(VI-2). HI parameter values are plotted in Fig. 2. The two quadrupole doublets for the M2 sites indicate that these are distinct categories of 6-fold coordinated sites with different local environments and HS(VI-1) involves less site-asymmetry than HS(VI-2), based on the QS values [37]. HS(VI-1) and HS(VIII) have relatively small average field $H_{hf}$ ~ 20 T values at low temperatures indicative of an unquenched orbital contribution, also discussed in the previous sub-section A.

In spite of the large volume change $\Delta V/V_0$ ~ 8% at the tetragonal-spinel/post-spinel structural transition the IS does not show an associated discontinuous decrease expected upon densification [44]. This is because the spinel sites undergo compensating changes to higher coordination HS(IV) $\rightarrow$ HS(VIII) which is expected to increase IS. The IS of the three components separate out in this pressure range as well, due to the different compressibilities of the M1 and M2 polyhedrons. $Fe^{2+}$ in the HS(VI-1)-designated site is apparently in a more compacted 6-fold coordinated local environment than HS(VI-2), inferred from its lower IS value. A further inference from the pressure dependences of the IS and QS parameters in Fig. 2 is that one of the octahedral sites HS(VI-2) with slightly increasing QS undergoes further distortion while the other sites HS(VIII) and HS(VI-1) become less distorted. These further pressure instigated distortions to HS(VI-2) will increase the crytal field (CF) splitting, with implications for spin crossover if a large enough CF splitting is attained [45, 46]. HS(VI-1) and HS(VIII) have monotonically decreasing QS values, perhaps indicating progressively decreasing polyhedral distortions as a cooperative lattice compensation for the behavior of HS(VI-2).

Therefore according to the MS data up to ~50 GPa three categories of sites have $Fe^{2+}$ parameters typical of the high-spin (HS) state, see Fig. 2 [47]. The LT spectra also show no evidence of a diamagnetic central QS doublet anticipated for low-spin (LS) $Fe^{2+}$ (atomic spin S=0). This contradicts the claims from XES experiments that spin crossover to a low-spin state is initiated as low as ~14 GPa [17]. Those experiments, limited to ~31 GPa, involve monitoring small intensity changes of relatively weak $K\beta'$ satellite features ascribed to the HS abundance and may be prone to systematic errors.

In our XRD data in the range 40−52 GPa the shift of the most intense diffraction peak near 2θ ~ 8° develops an appreciable increase in positive pressure dependence, see Fig 3(a). The refinements also show that the associated unit-cell volume of the *Cmcm* structure starts to deviate from the equation of state (EOS) of the HS post-spinel phase in this pressure range, see Fig. 4.

At pressures exceeding ~50 GPa changes are discerned in the RT and LT MS spectra compared to lower pressures. These changes are highlighted by vertical bars in the RT spectra at 55 GPa depicted in Fig.1 In the associated LT spectrum the smeared features in the central region at lower pressures now become more structured. When considering the profiles of the RT and LT Mössbauer spectra we are still able to fit each of these with a minimum of three sub-components, see Fig. 1. The spectral analysis shows that two categories of highly and less distorted 6-fold coordinated sites prevail. The spectral component associated with 8-fold coordinated sites has similar HI parameters to that of the preceding pressures below ~50 GPa. This is adequately modeled by a single broadened quadrupole doublet with QS ~ 1.5 mm $s^{-1}$. However the additional two doublets representing Fe in 6-fold coordination show starkly different temperature dependences. One of the remaining doublets with much larger QS ~ 2.3 mm $s^{-1}$ and IS ~0.7 mm $s^{-1}$ at RT and 55 GPa does show a magnetic splitting at LT, but with a much bigger $H_{hf}$ ~ 42 T compared to the preceding pressures of this *Cmcm* post-spinel regime. The other doublet with smaller QS ~ 1 mm $s^{-1}$ and IS ~ 0.4 mm $s^{-1}$ values at RT and 55 GPa, does not exhibit a magnetic splitting ($H_{hf}$ = 0 T) down to liquid helium temperatures and gives rise to intensity in a narrow region near zero velocity. This together with the evolution of large $H_{hf}$ ~ 42 T magnetic splittings for the other components accounts for the development of more structured features in the central regions of the spectra.

Fig. 5 also shows that the pressure dependence of the resistance at room temperature starts to decrease as the sample is compressed beyond ~40 GPa to a plateau at $P$ > 70 GPa.

Thus the combination of changes in, magnetic behavior seen in the (LT) magnetic Mössbauer spectra, pressure dependence of unit-cell volume and $R$(300 K) behavior onset in the vicinity of ~50 GPa, marks this as an electronic transition region.

The evolution of HI parameters across this phase transition region at ~50 GPa, Fig. 2, is thus interpreted in the following way. HS(VI-2) with large QS ~ 2.5 mm $s^{-1}$ and IS ~ 0.8 mm $s^{-1}$ at RT transitions to the component with much smaller QS ~ 1 mm $s^{-1}$ and IS ~ 0.4 mm $s^{-1}$ values and is diamagnetic. These are the tell-tale signatures of a high-spin to low-spin (HS $\rightarrow$ LS) transition [47]. The spin pairing accounts for the change to a relatively small QS value and the diamagnetism (atomic spin S=0) from the orbital population redistribution in $t_{2g}$- and $e_g$- derived sub-bands, $t_{2g}^4e_g^2$ (↑↓↑↑)(↑↑) $\rightarrow$ $t_{2g}^6e_g^0$ (↑↓↑↓↑↓)( ), see Fig 7. As HS(VI-2) only constitutes ~25% of 6-fold coordinated Fe sites the spin crossover in $Fe_2TiO_4$ only entails partial conversion of Fe sites to the LS state, namely, those octahedral locations with sufficiently large site-asymmetry corresponding to QS ~ 2.5 mm $s^{-1}$ in the HS state.

The $V(P)$ data of Fig. 4 exhibits a deviation from the HS *Cmcm* EOS onset in the range 40–45 GPa, leading to a unit-cell volume change of $\Delta V/V_0$ ~ 3.5%. This relatively small unit-cell volume change is considered corroboration that spin crossover is limited to a fraction of sites in the unit-cell, similar to what has been discerned in ferropericlase $Mg_{1-x}Fe_xO$ (for $x \leq 0.39$) [48]. The experimentally determined volume change is also compatible with an estimate obtained from considering the ~21% relative difference in octahedral volume $V_{oct} \propto (r_{Fe} + r_O)^3$ expected for HS and LS $FeO_6$ polyhedrons, where the radii in brackets are those of Fe and O nearest neighbor atoms [49, 50] ; and noting that spin crossover is limited to highly distorted octahedra (VI-2) constituting one quarter of the Fe sites (i.e., one sixth of all cations, [17]) in the structure.

Differences in onset pressures of LS signatures ascertained from MS, XRD and resistance data are attributable to different pressure transmitting media used and consequently how different degrees of hydrostaticity affect spin-crossover [51].

This partial spin-crossover has an associated cooperative lattice response. Less distorted HS(V1-1) sites with relatively small QS ~ 1 mm $s^{-1}$ values in the *Cmcm* phase at lower pressure now take on larger QS ~2.3 mm $s^{-1}$ values indicative of an increase in site-asymmetry [37], linked to the spin transition at the other VI-2 sites. The QS of HS(VI-1) does not exceed the critical QS value of ~2.5 mm $s^{-1}$ which was attained in HS(VI-2) and which signified sufficient octahedral distortion and an associated increase in CF to trigger the spin crossover at VI-2 sites.

Now consider the magnetic behavior deduced from fitting the LT spectra, see Fig. 1. The small $H_{hf}$ ~ 20 T values of HS(VIII) and HS(VI-1) of the *Cmcm* regime, give the smeared features in the centralized region of the magnetic spectrum, e.g., see LT spectrum at 40 GPa in Fig. 1. Above ~50 GPa these smeared contributions evolve to much sharper features, indicative of increased separation of the lines of each sextet spectral component due to a substantial increase in magnetic hyperfine splitting $H_{hf}$ ~ 42 T. This is close to the spin-only value contribution to the magnetic moment of $Fe^{2+}$ and its Fermi contact term contribution $H_C$ ~ 44 T. This signifies that the orbital moment term $H_L$ opposing $H_C$ has quenched at HS(VIII) and HS(VI-1) sites, recalling that $H_{hf} \sim H_C + H_L$. This is rationalized as follows.

At the onset of spin crossover at HS(VI-2) sites the cooperative lattice response of HS(VIII) and HS(VI-1) polyhedra is such that the $b_{2g}$ and $a_1$ lowest lying orbital levels are shifted to sufficiently below higher lying levels ($e_g$ and $a_2$, respectively) in Fig. 7, for orbital quenching to occur. The minority spin electron (↓) then predominantly populates these lowest lying orbitals which do not render an orbital moment contribution [43]. Furthermore the maximum value of the EFG, $V_{zz}$ , and thus QS [37] depends primarily on the 3*d* orbital occupancy of the minority spin electron. This is because $V_{zz} \propto e(2n_{xy} - n_{yz} - n_{xz} - 2n_{z^2} + 2n_{x^2-y^2}) / \langle r^3 \rangle$ , where $n_{xy}$ , $n_{yz}$ … denote the occupancy of the minority spin in orbitals |xy⟩, |yz⟩, …, respectively [38, 52]. The change in polyhedral distortion at both HS(VI-1) and HS(VIII), in response to the spin state change at HS(VI-2) and associated lattice response, is such that changes in low lying 3*d* level spacing in Fig. 7 lead to a change in orbital occupancy. Consequentially both the orbital moment quenches and $V_{zz}$ changes to result in the new appreciably bigger QS value of ~2.3 mm $s^{-1}$ for HS(VI-1). Note this is below the "critical" value of ~2.5 mm $s^{-1}$ which triggered spin crossover in HS(VI-2) in the *Cmcm* phase. Whereas in 8-fold coordinated HS(VIII) sites with a completely different level scheme to 6-fold coordinated sites (see Fig. 7), this change in orbital occupancy has much less impact on $V_{zz}$ than in 6-fold coordinated sites, but still leads to orbital moment quenching [43].

### C. 55–90 GPa, persistent partial spin crossover and *Pmma* post-spinel phase

When considering the profiles of the RT and LT Mössbauer spectra upon further compression beyond ~55 GPa, we are still able to fit each of these with a minimum of three sub-components, see Fig. 1. The analysis of the MS spectra show the abundances of the three components are in the ratio 1:0.5:0.5. The component representing Fe in 8-fold coordinated sites being of equal abundance to the combination of the other two 6-fold coordinated sites. The IS and QS parameters of the components have monotonically decreasing values upon further compression steps to ~90 GPa. The partial spin conversion initiated in the *Cmcm* phase remains limited to a fraction of the Fe sites up the highest pressure of ~90 GPa attained in this study. This is exemplified in the magnetic spectrum at the highest pressure of ~89 GPa in Fig.1. A doublet of ~25 % abundance occurs in a narrow region near zero velocity corresponding to the diamagnetic LS species. The remainder of HS(VI-1) and HS(VIII) sites continue to both have a magnetic ground state involving a quenched orbital moment and associated $H_{hf}$ ~ 42 T.

Yamanaka *et al.* [17] contend from x-ray pressure studies that a further structural adjustment occurs in the vicinity of ~50 GPa. They have refined the patterns above ~50 GPa in terms of the *Pmma* space group, a non-isomorphic sub-group of *Cmcm*. Yamanka *et al.* only show two data points for such *Pmma* refinements in the range 50–60 GPa. It may be noted that in our XRD data, extended to much higher pressures of 86 GPa, conspicuous splittings occur in some of the reflections in the range 52–56 GPa indicative of some symmetry lowering, see Fig. 3. In accord with Yamanaka *et al.* [17], we have chosen to refine the patterns above ~50 GPa in terms of the symmetry-lowering sub-group *Pmma* of *Cmcm*. A slightly new trend in the pressure dependence of the volume is then followed up to at least 80 GPa, see Fig.4. Yamanaka *et al.* have attributed this structural adjustment to atomic ordering of Fe/Ti in the 6-fold coordinated sites. Within this post-spinel variant, two distinct Fe sites (sets of bond lengths) occur in both 8-fold and 6-fold coordination environments.

Recall that spin-state changes at specific octahedral sites (VI-2) are onset within the post-spinel *Cmcm* structure preceding this transition to the *Pmma* phase. We speculate that atomic diffusion processes are enhanced for the resultant LS species with smaller atomic volume. This facilitates the subsequent *Cmcm* → *Pmma* disorder/order transition in the sub-lattice of 6-fold coordinated sites.

The temperature dependence of the resistance measured to a highest pressure of ~68 GPa indicates continued activated transport processes (linearity in ln($R$) vs 1/$T$ ) similar to the preceding *Cmcm* structural regime. The charge gap ($E_G = 2E_a$) deduced from fits to the linearized data in the inset of Fig. 5 is $E_G$ ~120 meV at ~68 GPa. The pressure dependence of the resistance at RT to much high pressure shows a plateau behavior from ~68 GPa onwards up to ~80 GPa. This suggests that the sample remains non-metallic to the highest pressure of this study, i.e., to near a megabar.

## IV. Summary and conclusions

We arrive at the following conclusions on the pressure response of the magnetic-electronic behavior of ulvöspinel $Fe_2TiO_4$ and its relation to structural transformation and electrical-transport behavior :

(i) The cubic-to-tetragonal transition rises from $T_{J\text{-}T}$ ~ 163 K to room temperature by ~ 8 GPa, after which tetragonality at RT prevails until ~16 GPa when a new structural transition is onset. Complex magnetic spectra occur , reflecting levels of disorder in the Fe/Ti *B* sub-lattice. Throughout this pressure range there is a large average $H_{hf}$ distribution component designated HS(VI-2), corresponding to a certain degree of cation segregation and highly asymmetric Fe local environments (large QS). In addition two different smaller average $H_{hf}$ distribution components also occur. These are related to a different degree of Fe/Ti segregation at HS(VI-1)-designated 6-fold coordinated sites and to 4-fold coordinated sites designated HS(IV).

(ii) The tetragonal spinel transforms to the orthorhombic post-spinel in the range 16–20 GPa as discerned in the precipitous ~8% decrease in unit-cell volume and decrement by two orders of magnitude of the room-temperature resistance. The post-spinel is identified to be of the $CaTi_2O_4$-type with space group *Cmcm*, in accord with previous work. This seems to only have had minor effects on the hyperfine interaction parameters and magnetic behavior across the structural transition. Small average $H_{hf}$ (HS(VI-1)) and large average $H_{hf}$ (HS(VI-2)) categories of 6-fold coordinated sites prevail. Tetrahedral sites transform to higher (8-fold) coordinated, HS(VIII)-designated, sites with a small average $H_{hf}$ and other similar hyperfine interaction parameters to HS(IV) sites. All of this perpetuates to 40–45 GPa when a change to a steeper pressure dependence of unit-cell volume occurs. This leads to a ~3.5% change in unit cell volume and $P$ ~ 50 GPa is identified as another phase transition region.

(iii) In the vicinity of ~50 GPa changes are discerned in RT and LT MS spectra compared to those typical of the HS state of the *Cmma* phase. Magnetic spectra down to liquid helium temperatures comprise large average $H_{hf}$ ~ 42 T contributions and a diamagnetic central doublet. This doublet is maximally ~25% of the Fe sites from the analysis of spectra collected at higher pressures. The HI parameters of the doublet have IS and QS values typical of a low-spin state. This evolves from the highly distorted HS(VI-2) sites and comprises about half of the 6-fold coordinated Fe sites in the lattice. The rest of the Fe locations have large average $H_{hf}$ ~ 42 T values which signify quenching of orbital moment contributions at HS(VI-1) and HS(VIII) sites. These findings are in contrast to previous claims of spin crossover initiating as low as ~14 GPa [17]. Indeed spin crossover does occur but only at specific 6-fold coordinated sites and it initiates beyond 40 GPa within the *Cmcm* phase. The relatively small volume change of ~3.5% across the spin crossover regime is considered corroboration of only partial conversion of 6-fold coordinated Fe sites to the low-spin state.

(iv) At pressures beyond ~55 GPa XRD profiles indicate symmetry lowering to the *Pmma* sub-group supposed to be associated with Fe/Ti order at 6-fold coordinated sites. Partial spin crossover initiated in the preceding *Cmcm* phase facilitates a subsequent disorder/order transition to the *Pmma* phase in the vicinity of ~55 GPa. Analysis of the MS spectra indicates that the partial spin crossover, limited to the highly distorted 6-fold coordinated sites of ~25 % abundance, persists up to ~90 GPa in the *Pmma* phase.

(v) Arrhenius activated nearest-neighbor hopping is applicable to the compound across all structural transition boundaries. The room temperature resistance tends to a plateau at pressures beyond ~55 GPa. The charge gap is inferred to be ~100 meV in the range 80–90 GPa, based on the behavior of the resistance at room temperature and pressure dependence of activation energies obtained at lower pressures.

Thus the very high pressure post-spinel phase of $Fe_2TiO_4$ at 55–90 GPa is non-metallic, is stabilized in the *Pmma* structure involving Fe/Ti ordered in 6-fold coordinated sites and has a partially spin-converted lattice involving LS states at ~25% of the Fe sites. The remainder of the 6- and 8-fold coordinated Fe sites have magnetic ground states in which the orbital moment is quenched, to yield spin–only internal magnetic field values of ~42 T.

Our XRD pressure measurements to above 80 GPa, well beyond the maximum pressure of previous studies, indicate that that there is a ~33% reduction in unit-cell volume to these extremes. Yet the presumably appreciable 3*d* band-broadening associated with such densification is not sufficient to disrupt strong electron correlations and close the charge gap. The effective on-site repulsion parameter ($U_{eff}$ determining the gap between upper and lower Hubbard bands), renders the non-metallic (Mott insulator type) behavior to these partially filled 3*d* band oxides. In the high-spin $d^6$($Fe^{2+}$) electronic configuration $U_{eff}$ is constant as a function of pressure and increases at spin crossover in 6-fold coordinated sites [46]. Moreover the high-pressure *Pmma* structure involves ordered Fe and Ti at 6-fold coordinated sites. $Ti^{4+}$ has an empty 3*d* band and should therefore be able to accommodate a charge carrier in a delocalization process, because there would be no energy cost $U_{eff}$ from double-occupation at Ti sites. We therefore infer that it is the *Cmcm* (disorder) → *Pmma* (order) transition and resultant Fe/Ti ordered configuration which does not render sufficient Fe-Ti orbital overlap for delocalization to occur. In addition the disposition throughout the lattice of low-spin sites with increased $U_{eff}$ values in only one quarter of the Fe sites is sufficient to ensure that the cost of double-occupation (onsite Coulomb repulsion) continues to supersede the kinetic energy gained in band broadening and associated delocalization. As such strong electron correlations prevail in the atomically ordered *Pmma* phase of non-metallic $Fe_2TiO_4$, in spite of appreciable 3*d* band broadening at high density.

## Acknowledgements

We acknowledge with gratitude receipt of the $Fe_2TiO_4$ sample from X. Wu (School of Earth and Space Sciences, Peking University). We also acknowledge professional assistance from Rudolph Rüffer, Aleksandr Chumakov and support staff during the synchrotron Mössbauer source measurements at the ID-18 beamline of the ESRF. This research was supported by the Israel Science Foundation (Grant No. 1189/14). G.R.H. acknowledges financial support from the National Research Foundation of South Africa (Grant No. 105870).

**Figure captions**

FIG. 1. Representative Fe Mössbauer spectra at room temperature (RT, left panel ) and low temperatures (LT, right panel), in the pressure regimes discussed in the text. Solid line through the data points represent the overall fit to the data from the sum of sub-components shown. Dotted line sub-components refer to the 4-fold or 8-fold coordinated sites and other sub-components refer to 6-fold coordinated sites, discussed in the text. The vertical lines are a guide to the eye as to where changes occur in the spectral profile compared to the preceding pressure regime. The shaded component in the *Pmma* phase represents low-spin Fe sites.

FIG. 2. Pressure dependences of the hyperfine interaction parameters from best fits to the Mössbauer spectra. Vertical error bars do not exceed twice the size of the symbols but are omitted to avoid clutter in the plots. Top panel has saturation magnetic hyperfine field $H_{hf}$ values from the low temperature (LT) spectra. QS and IS parameters from spectra at room temperature (RT) are in the middle and bottom panels. Three pressure regimes demarcated by the vertical lines have been identified and discussed in the text. Roman numerals IV, VI and VIII refer to 4- 6- and 8-fold coordinated tetrahedral, octahedral and bicapped trigonal prismatic sites, respectively. Spectra in the ambient–20 GPa range, where XRD discerns cubic and then tetragonal behavior at 300 K, are best fit at this temperature with three QS components having similar IS values as discussed in the text. These are associated with two categories of 6-fold coordinated Fe sites, HS(VI-1) and HS(VI-2), and 4-fold coordinated sites HS(IV). Spectra in the regime 20-50 GPa are deconvoluted in a similar way. The structural transformation of 4-fold to 8-fold coordinated HS(VIII) sites and evolution of the 6-fold coordinated sites across the $I4_1/amd$ → *Cmcm* reconstructive transition are described in ref. [18]. Spectral analysis suggests two categories of large QS (HS(VI-2)) and small QS (HS(VI-1)) persist into this regime. Large $H_{\mathrm{hf}}$ values associated with orbital moment quenching are delineated by the (omq) label. Parameters for the low-spin sites are delineated by the (LS) labels.

FIG. 3. (a) Pressure evolution of the XRD patterns up to ~86 GPa ($\lambda$ = 0.3738 Å). Contamination reflections from the rhenium gasket are indicated. The dashed box

highlights where changes in reflection profiles occur at very high pressure, supposed to be an indicator of a *Cmcm* → *Pmma* symmetry lowering due to a Fe/Ti disorder → order transition at 6-fold coordinated sites, as discussed in the text. (b) Examples of refinements of the XRD patterns at pressure in the various structural phases.

FIG. 4. Pressure dependences of the unit-cell volume of the various structural phases. The vertical error bars do not exceed the size of the symbols. Solid lines through the data symbols are fits with a Birch-Murnaghan equation of state (EOS) to obtain bulk modulus $K_0$ and constraining its derivative at $K_0'$=4 [53]. The vertical dashed lines demarcate the pressures where structural transitions are onset.

FIG. 5. Pressure dependence of the resistance at 300 K in the main panel. Different symbols refer to various four-probe arrangements and inter-electrode distances, based on selections from the six electrodes configured in the sample cavity. The pressure dependences are the same in all three cases, as a check of any effects of pressure distribution or electrode-sample contact issues. Top right inset shows linearized temperature dependent data at various pressures assuming Arrhenius activated hopping transport, $\ln(R) \propto \frac{E_a}{k_B T} + const.$ . The activation energy $E_a$ is obtained from the slope of fits to these plots, from whence the charge gap for intrinsic conduction $E_g = 2E_a$ is derived. Bottom left-hand inset shows the pressure dependence of $E_a$ ; the solid line through the data points is to guide the eye. Dashed vertical lines in the main panel delineate onset pressures to new structural phases deduced from appreciable changes in pressure dependences of $R$(300 K) and $E_a$ for comparison with XRD experiments, see text and preceding Fig. 4.

FIG. 6. High-pressure structure of $Fe_2TiO_4$ post-spinel with space group *Cmcm*. Locations of 6-fold and 8-fold coordinated sites are indicated. Fe and Ti are disordered on the 6-fold coordinated sites. The left panel is a perspective view and right panel shows the projection down the *c*-axis and characteristic bi-directional herring bone arrangement.

Fig. 7. Schematic of crystal-field (CF) split 3*d* electronic level scheme for $Fe^{2+}$ involving tetragonal distortions in octahedral (6-fold) coordination and compression of bicapped trigonal prism (8-fold coordination site), the latter is adapted from Burdett *et al.* [54]. Irreducible representation symmetry labels for the orbitals are indicated in the yellow blocks.

**Figure 1   Xu *et al.***

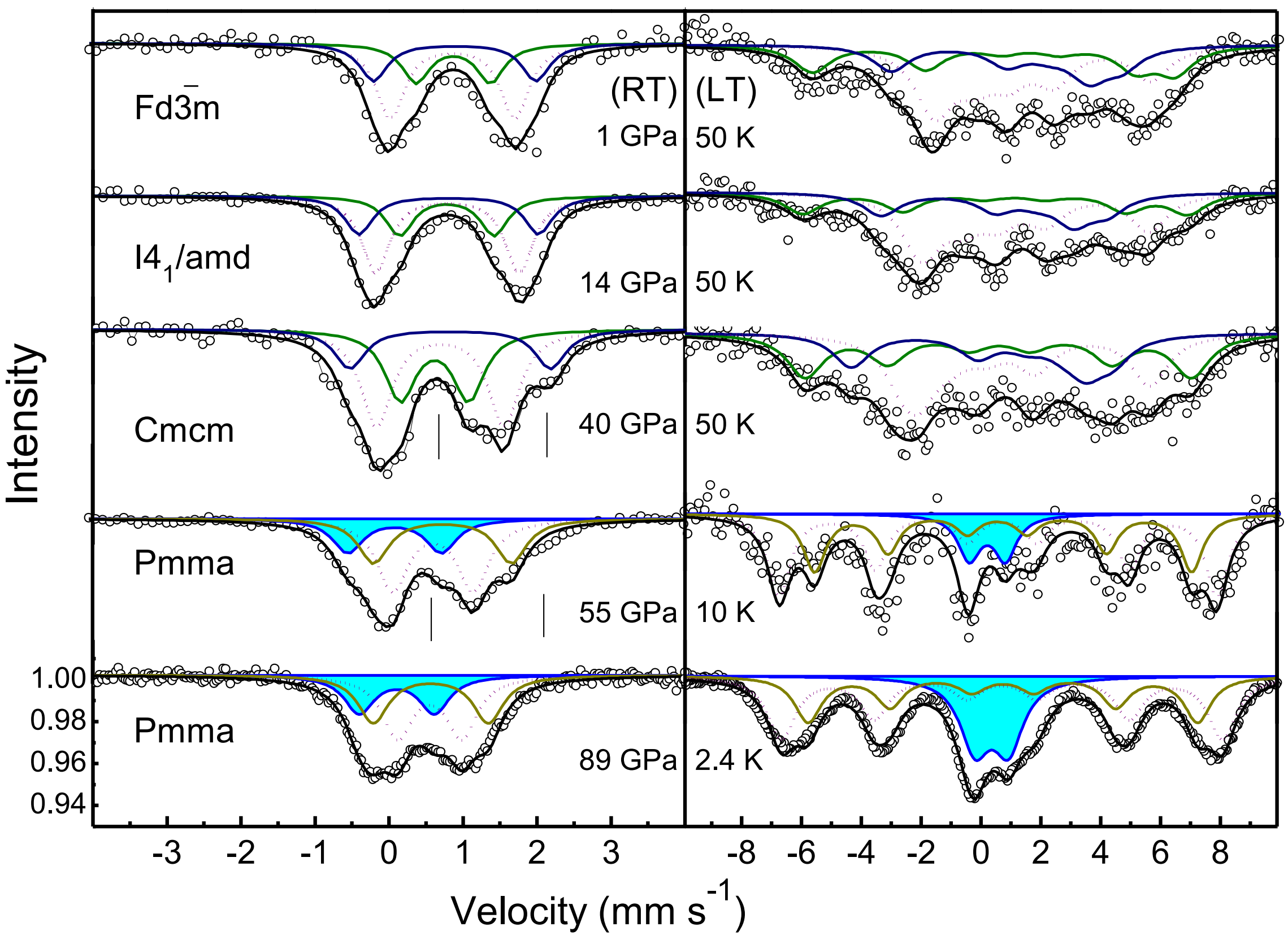

**Figure 2 Xu *et al.***

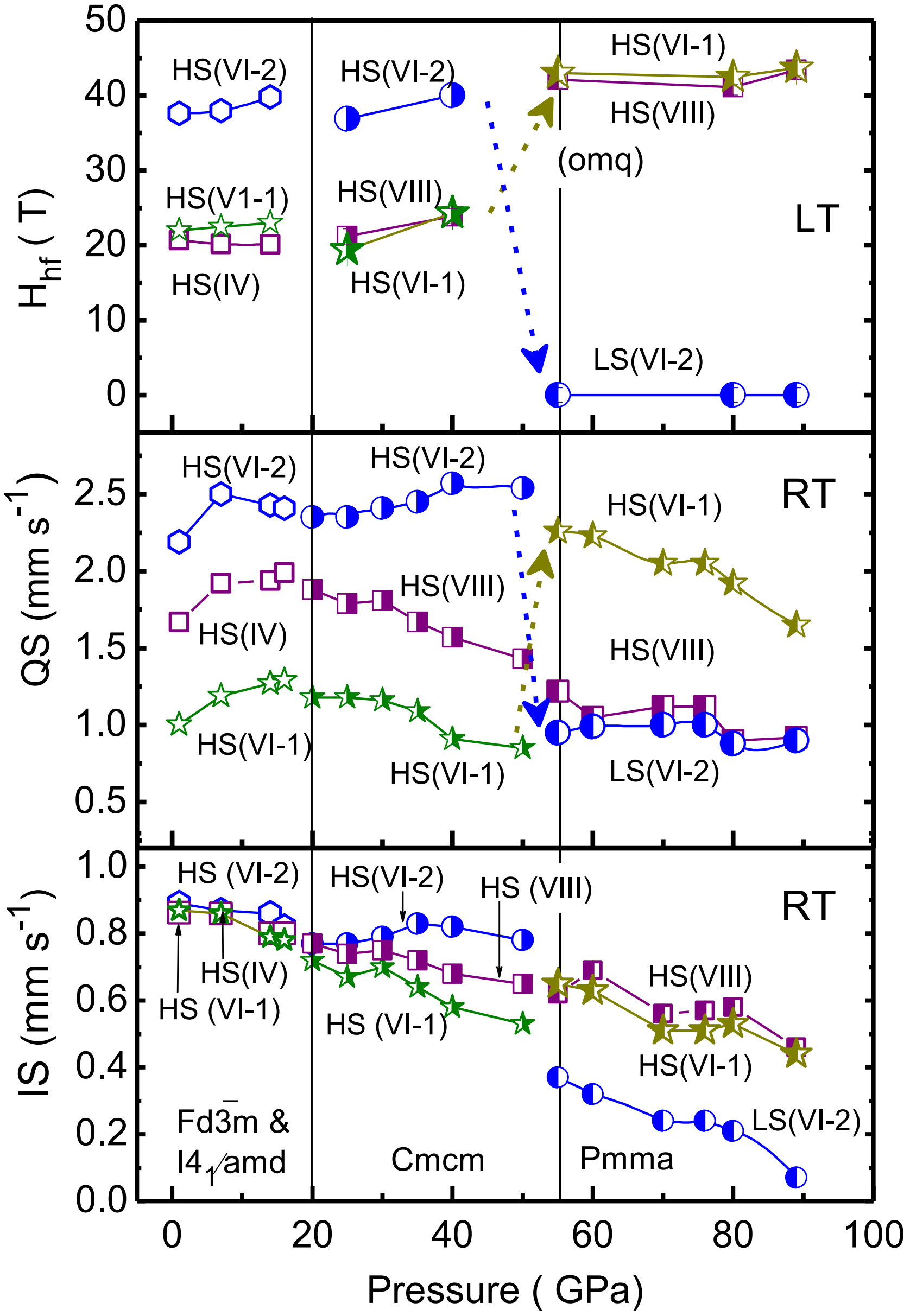

**Figure 3   Xu *et al.***

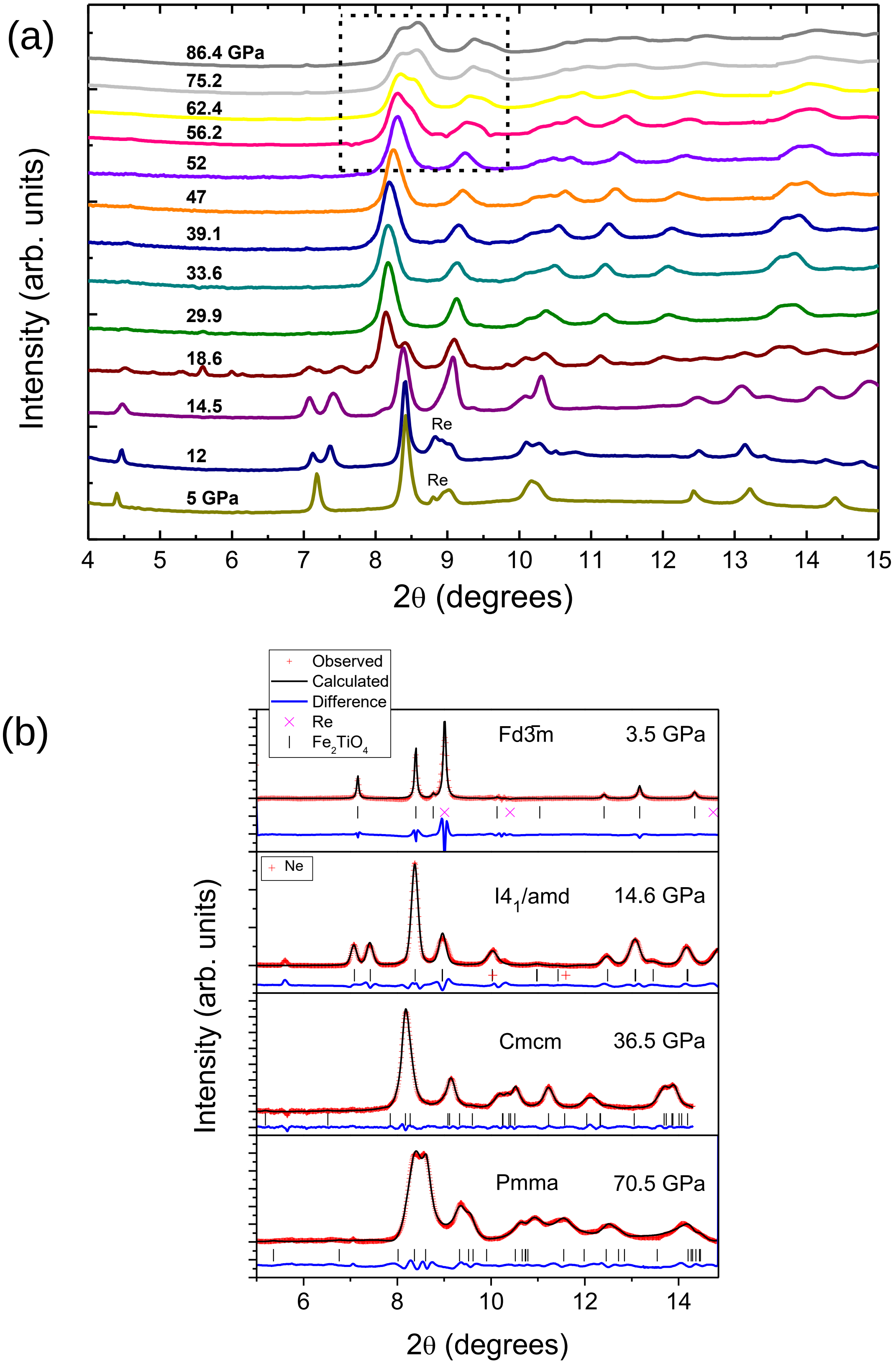

**Figure 4 Xu *et al.***

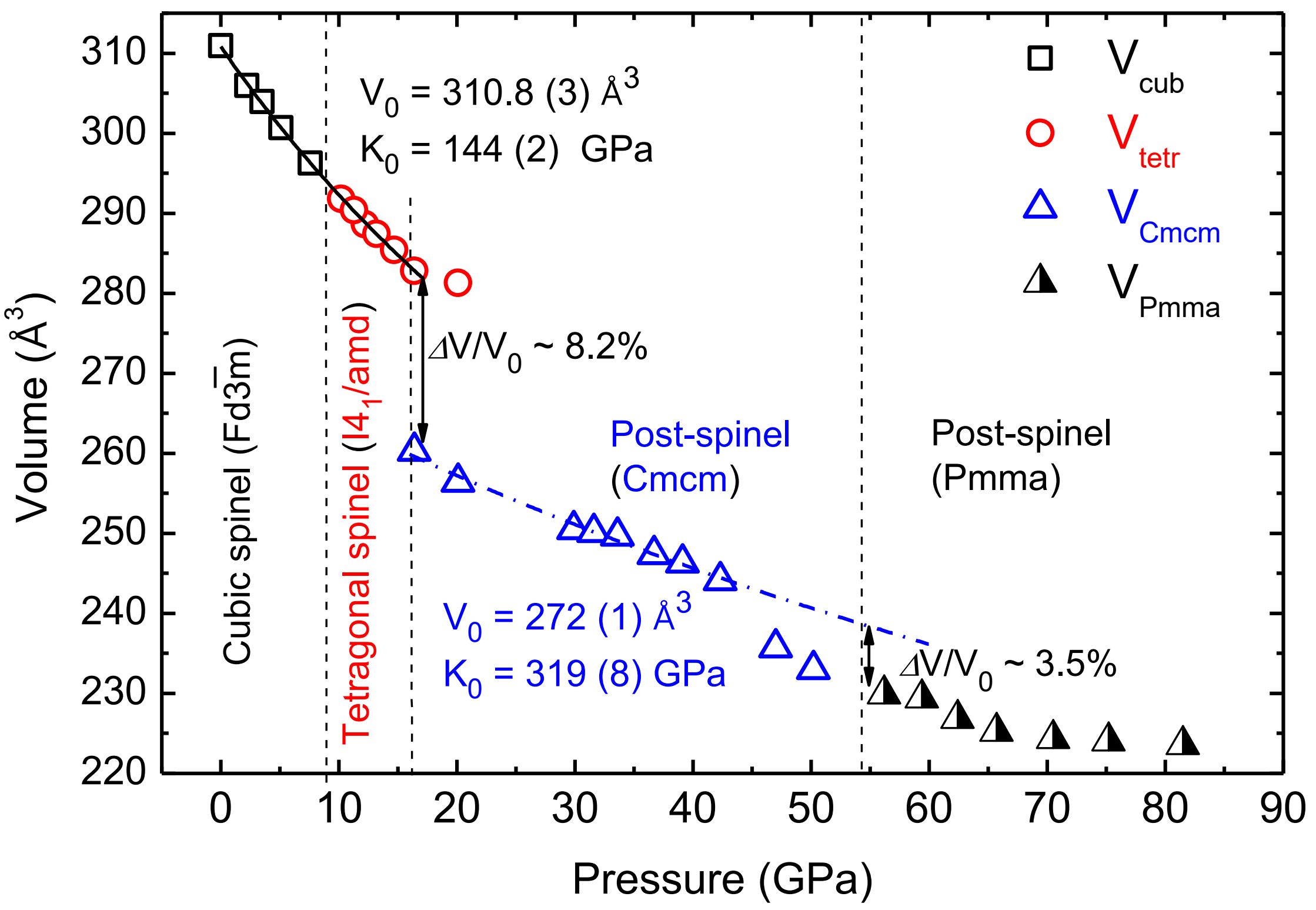

**Figure 5 Xu *et al.***

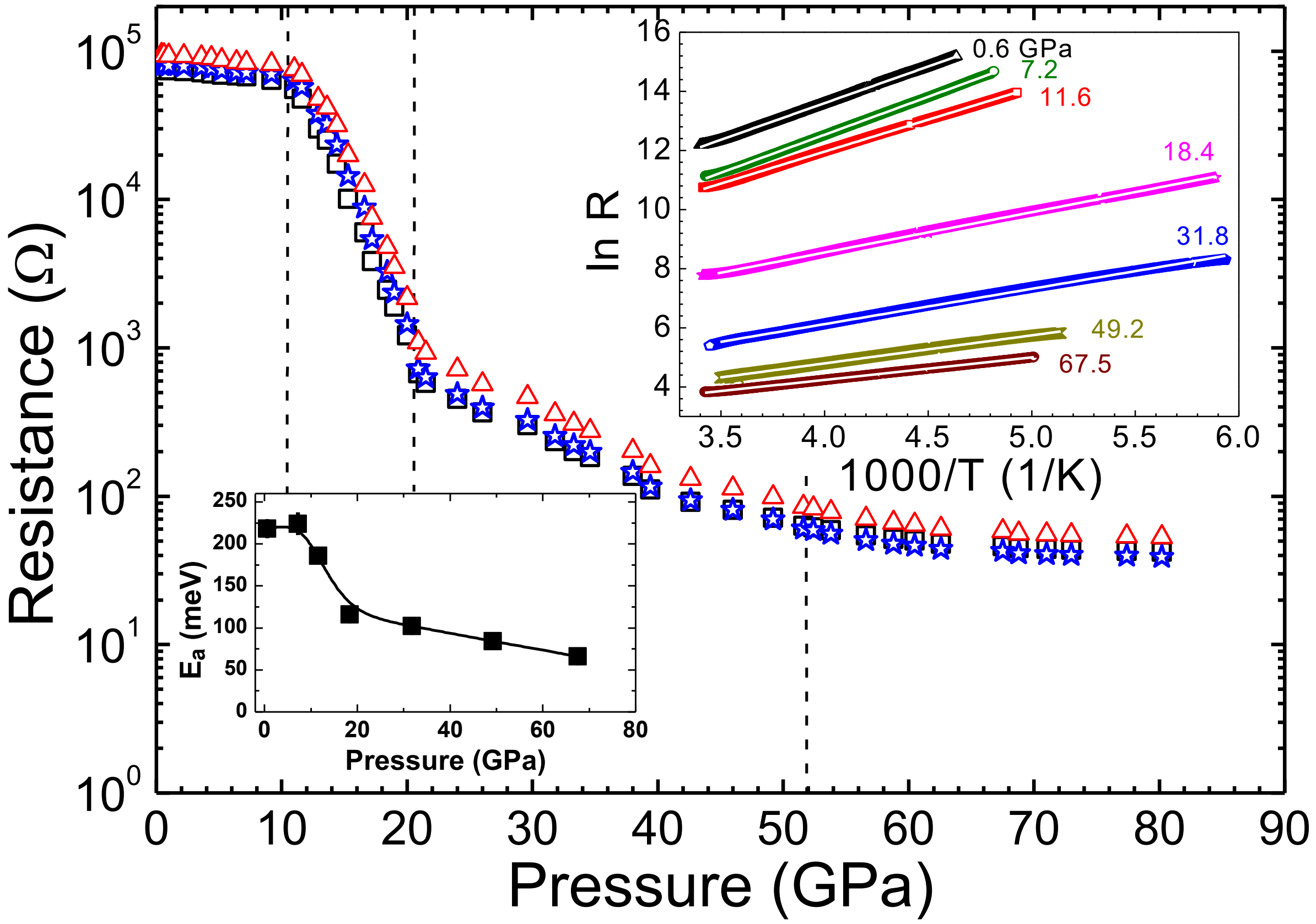

**Figure 6 Xu *et al.***

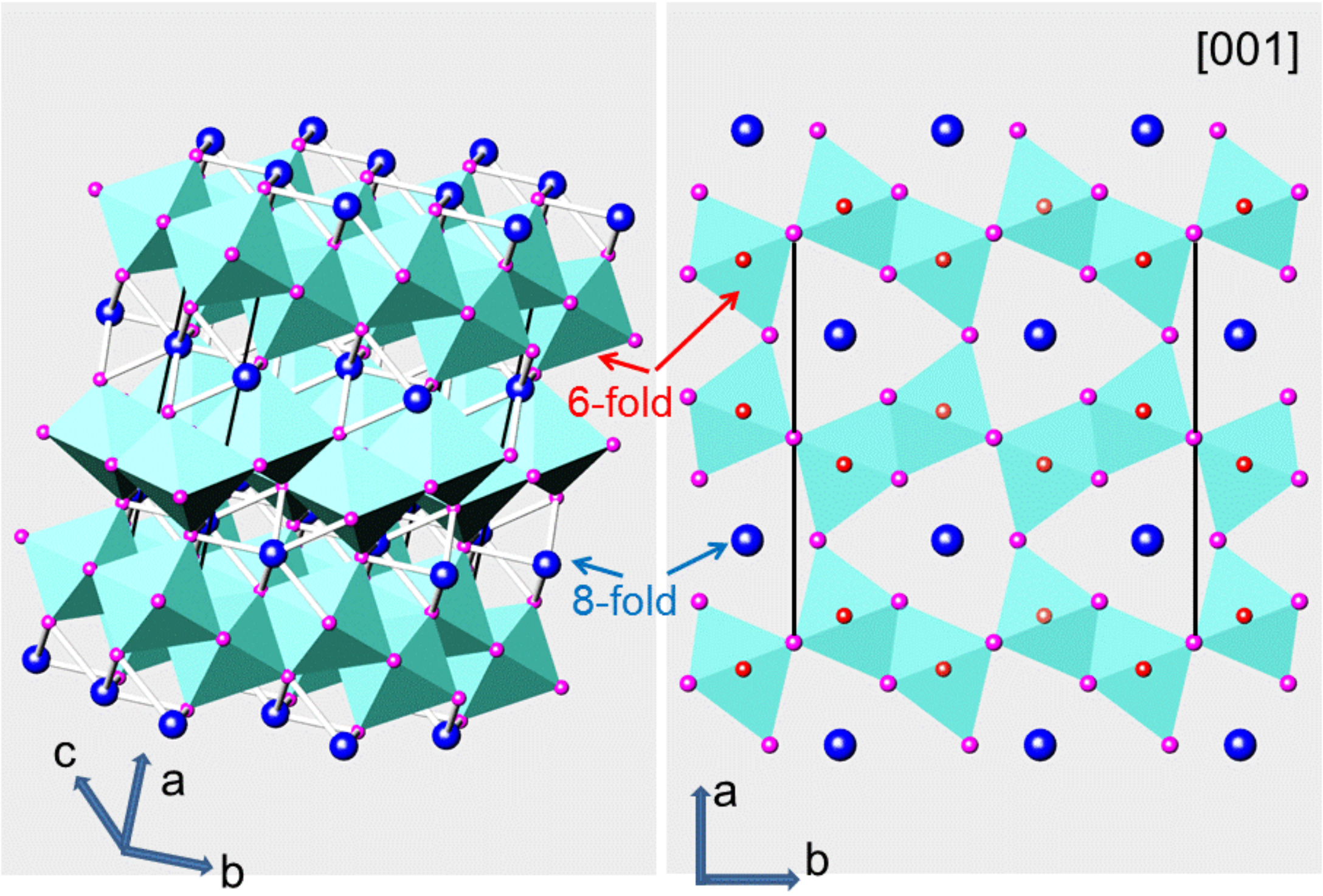

**Figure 7 Xu *et al.***

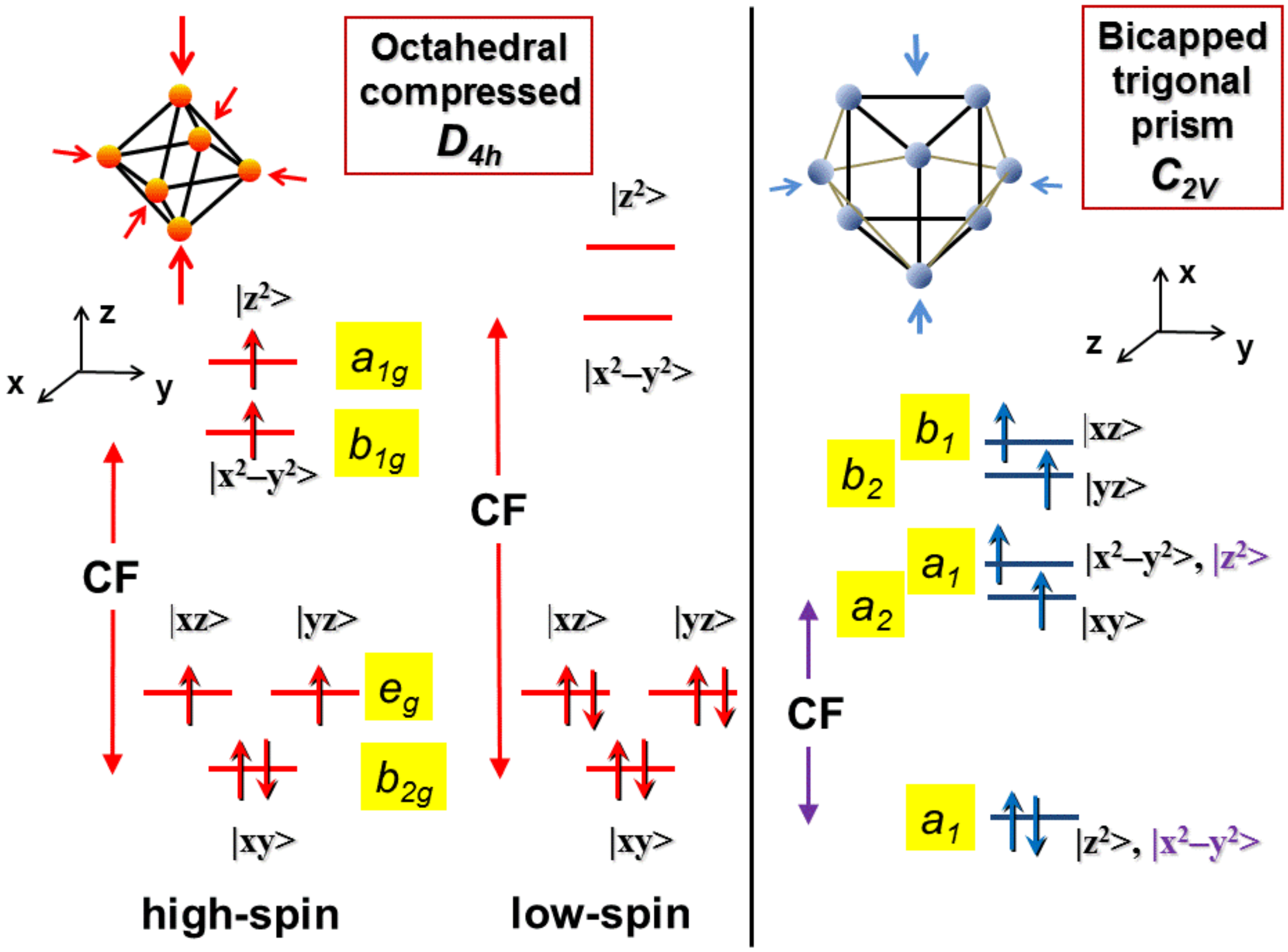